\documentclass[conference]{IEEEtran}
\IEEEoverridecommandlockouts
\usepackage{cite}
\usepackage{amsmath,amssymb,amsfonts}
\usepackage{algorithmic}
\usepackage{graphicx}
\usepackage{textcomp}
\usepackage{xcolor}
\usepackage{comment}

\usepackage{hyperref} 
\hypersetup{ colorlinks=true, linkcolor=black, filecolor=black, urlcolor=cyan, }

\def\BibTeX{{\rm B\kern-.05em{\sc i\kern-.025em b}\kern-.08em
    T\kern-.1667em\lower.7ex\hbox{E}\kern-.125emX}}
\usepackage{fancyhdr}
\usepackage{kantlipsum}
\fancypagestyle{plain}{
\fancyhf{}
\fancyhead[C]{Conference on \LaTeX} 

}
\usepackage{eso-pic}
\begin{document}
\AddToShipoutPictureBG*{
\AtPageUpperLeft{
\setlength\unitlength{1in}
\hspace*{\dimexpr0.5\paperwidth\relax}
\makebox(0,-0.75)[c]{\textbf{2022 IEEE/ACM International Conference on Advances in Social Networks Analysis and Mining (ASONAM)}}}}

\title{Understanding the Impact of Culture in Assessing Helpfulness of Online Reviews
}

\author{\IEEEauthorblockN{Khaled Alanezi}
\IEEEauthorblockA{\textit{Computer Department} \\
\textit{College of Basic Education, PAAET,  Kuwait}\\
kaa.alanezi@paaet.edu.kw}
\and
\IEEEauthorblockN{Nuha Albadi, Omar Hammad, Maram Kurdi and Shivakant Mishra }
\IEEEauthorblockA{\textit{Department of Computer Science} \\
\textit{University of Colorado, Boulder, CO, USA}\\
{\{Nuha.Albadi$|$omha6093$|$Maram.Kurdi$|$mishras\}@colorado.edu}
}
}

\maketitle
\IEEEoverridecommandlockouts
\IEEEpubid{\parbox{\columnwidth}{\vspace{8pt}
\makebox[\columnwidth][t]{IEEE/ACM ASONAM 2022, November 10-13, 2022}
\makebox[\columnwidth][t]{978-1-6654-5661-6/22/\$31.00~\copyright\space2022 IEEE} \hfill} \hspace{\columnsep}\makebox[\columnwidth]{}}
\IEEEpubidadjcol
\begin{abstract}
Online reviews have become essential for users to make informed decisions in everyday tasks ranging from planning summer vacations to purchasing groceries and making financial investments. A key problem in using online reviews is the overabundance of online that overwhelms the users. As a result, recommendation systems for providing helpfulness of reviews are being developed. This paper argues that cultural background is an important feature that impacts the nature of a review written by the user, and must be considered as a feature in assessing the helpfulness of online reviews. The paper provides an in-depth study of differences in online reviews written by users from different cultural backgrounds and how incorporating culture as a feature can lead to better review helpfulness recommendations. In particular, we analyze online reviews originating from two distinct cultural spheres, namely Arabic and Western cultures, for two different products, hotels and books. Our analysis demonstrates that the nature of reviews written by users differs based on their cultural backgrounds and that this difference varies based on the specific product being reviewed. Finally, we have developed six different review helpfulness recommendation models that demonstrate that taking culture into account leads to better recommendations.
\end{abstract}



\section{Introduction}
\label{sec:introduction}

Online reviews play a crucial role in every user's decision-making process, ranging from vacation planning, purchasing groceries, researching the next new automobile, summer reading plans, and so on. Indeed, 
allowing users to write online product reviews has become a necessary feature in almost every e-commerce website. Unfortunately, because of the globalized economy and increased travel, users of these websites are increasingly confronted with an overabundance of product reviews written by users from different cultural backgrounds, which overwhelms them from making appropriate decisions
Indeed, to help the user digest the overwhelming information provided by these often-large numbers of reviews, e-commerce websites are opting for automatic translation to address the issues of the language barrier. In addition, comments combined with star ratings are also rated for {\it helpfulness} to allow users to skip through unnecessary reviews that don't provide any added value. In this paper, we argue that all these features are undoubtedly necessary but still are inadequate in providing website users with a complete understanding of the product being reviewed. 
In particular, culture is known to be a massive influencer of consumer behavior \cite{mccort1993culture,shavitt2016culture,otnes2012gender} and writing an online review is not an exception. Hence, we propose that platforms providing review helpfulness recommendations must utilize culture as an essential input in their recommendations.

We provide an in-depth study of differences in online reviews written by users from different cultural backgrounds and how incorporating culture as a feature can lead to better review helpfulness recommendations. First, we analyze online reviews from two distinct cultural spheres, namely Arabic and Western cultures. The main reason for considering these two cultures is the users with backgrounds in these two cultures comprise a relatively large chunk of Internet users, thereby helping inform similar studies covering other cultures. Further, to understand how the impact of culture varies depending on specific product types and perceive review helpfulness, we analyze two types of products, namely hotels and books. We chose these two types of products because users heavily rely on electronic word-of-mouth (eWOM) when buying or subscribing to these two products. 

This paper makes the following important contributions: 
\begin{itemize}
    \item We show that the users' cultural background plays an important role in shaping the reviews they write, and the differences in the reviews by users from different cultural backgrounds are statistically significant.
    \item We show that the impact of the users' cultural background on the reviews they write varies depending on the product being reviewed, and this difference is statistically significant.
    \item We develop review helpfulness recommendations that take culture into account and show that this leads to better recommendations.
    \item We make public two manually labeled datasets for review helpfulness in the Arabic language \footnote{\url{https://osf.io/n3psk/?view\_only=319598f17b5942f6b12038b09309ef79}}
\end{itemize}

\section{Background and Related Work}
\label{sec:Related}


\subsection{Arab Culture}
Arab countries span the Middle East and North African region (MENA). While Arab culture varies across MENA countries, Arabs in general tend to be more collectivist rather than individualist \cite{ourfali2015comparison}. In collectivistic culture, group solidarity and meeting social expectations are prioritized over individual gains \cite{semaan2017impression}. Middle Eastern societies value loyalty, honor, respect for older individuals and religion \cite{grafton2012arab}. In fact, religion plays an integral part of Arabs day to day communications with around 93\% of Arabs practicing Islam \cite{pew2015religious}. Thus, privacy and modesty, which mainly stem from Quran, are highly valued in Arab culture, particularly for women \cite{abokhodair2016privacy}. These cultural values can potentially impact the way people communicate online, adopt technology, provide consumer reviews, and so on.  

\subsection{Cross-Cultural Product Reviews}
Understanding cultural differences and its impact on technology adoption has been of interest to many HCI/CSCW scholars \cite{cheng2021country,li2020studying, choi2016cross}. While we are not aware of any work that has studied the impact of Arab and/or Muslim culture on online consumer product reviews, some notable research has been done to investigate Middle Eastern socio-cultural differences in other online contexts such as mitigating domestic abuse \cite{rabaan2021daughters}, seeking a potential spouse  \cite{al2021saudi}, sharing Quran verses \cite{abokhodair2020holy}, and using bots for political reasons \cite{albadi2019hateful}. 

The impact of culture on online consumer product reviews has been investigated in other different cultures. For instance, the authors in \cite{park2009antecedents} compared how online reviews and perceived helpfulness affect the purchase decision of Korean consumers versus American consumers. In another comparative study \cite{lin2018culturally}, the authors studied differences in reviewing behavior and the relationship between reviews and culture and how these differences change over time using data from two culturally different marketplaces, i.e., Amazon U.S. and Amazon Japan. The influence of culture and the differences in reviewing behavior between Chinese and U.S. costumers have been investigated in multiple online settings such as hotels \cite{kim2018effects}, gaming \cite{tsang2009does}, movies \cite{koh2010online}, and e-commerce \cite{zhu2017understanding}. 

In a multi-cultural study \cite{hong2016culture}, researchers found that reviewers from collectivistic cultures are more likely to conform to prior ratings and opinions and are less likely to express emotions in their reviews compared to reviewers from individualistic cultures. The study by \cite{kim2018effects} suggested that a culturally-oriented review information would be highly valuable to costumers to help them make an informed purchase decision.

\subsection{Review Helpfulness}
The power of product reviews has a huge impact on customers decisions. About 93\% of the customers had their purchase decisions influenced by reading products reviews \cite{CUSTOMER_EXPERIENCE}. Thus, a significant body of work has investigated the helpfulness of reviews in different aspects.  

Using a machine learning model, the authors in \cite{kim2006automatically} were able to automatically predict the helpfulness of online product reviews. Using Amazon.com dataset, they trained an SVM regression model in order to rank product reviews as helpful or not. Their model achieved a scores of 0.656 in Spearman correlation coefficient. In addition, they reported that the most important features yielded by their regression model are the review length, its uni-grams and its product rating. 

Another study that was conducted by \cite{malik2017helpfulness} has predicted review helpfulness by building a deep neural network classification model and investigated the role of positive and negative emotions towards review helpfulness prediction. They trained six different classification models on combination of a hybrid set of extracted features from online products reviews. These features includes product, reviewer, visibility, linguistic, readability, sentiment, and product features. Other features that they extracted were discrete positive emotions and negative features. The best prediction result (0.89 $F_{1}$) was achieved by training a deep neural network classification model using hybrid set of features along with discrete positive emotions feature. 

The researchers in \cite{liu2015makes} focused on two elements when predicting reviews helpfulness, the review itself and its writer. Based on their textual regression model, they found that a significant impact on review helpfulness was associated with reviewers' identity disclosure. Moreover, there was an association between positive reviews and the perception of helpfulness. Review readability was a strong indicator of review helpfulness perception.

\section{Analysis}
\label{sec:analysis}
\subsection{Data Set}
\label{sec:data-set}

In order to gain a full understanding of the impact of users' cultural backgrounds while writing online reviews, we begin by analyzing how users from the two cultures under study review products online. We included multiple review datasets, each of which contained reviews of star ratings and corresponding textual reviews. Past research has shown that both types of feedback (star rating and textual review) impact a user's choice when selecting products online \cite{devedi2017study,ghose2007designing,zhang2010impact}. To study the nature of reviews, we processed review text to study patterns related to the length of reviews, sentiments expressed and any product aspects mentioned. In addition to cultural differences, we believe that the type of product being reviewed also impacts the way users write a review and provide associated star ratings. Therefore, our analysis looked at reviews of two different products, namely hotels and books. In summary, the analysis looked at four different datasets to cover the two cultures and two products under analysis. These include a review of hotels in Arabic~\cite{elnagar2018hotel}, a review of hotels in English~\cite{english-hotels-ds}, a review of books in Arabic~\cite{aly2013labr} and a review of books in English~\cite{wan2018item}. 
{The datasets came from two different platforms: Booking.com and Goodreads, both of which are internationally recognized as the most popular sites for hotels and books respectively. These sites cover a very wide range of hotels and books respectively, which along with the sheer size of the datasets ensures that we have a fairly complete coverage of  both cultures along with variations in hotel quality and book categories}. Table \ref{table:summary} provides a summary of these four datasets with associated general statistics. For Arabic hotel, English hotel, and Arabic books review analysis, we included all the reviews in the corresponding datasets. However, the English books review dataset included a significantly large number of reviews compared to the other three datasets (300K-500K vs. 15M reviews). Due to the sheer size of the English books review dataset, we picked 330K reviews randomly using the reservoir sampling for our analysis. 

\begin{table}
  \caption{General Stats of Used Data Sets}
  \label{table:summary}
  \begin{tabular}{|l|c|c|c|c|}
    \hline
     & \multicolumn{2}{|c|}{Arabic} & \multicolumn{2}{|c|}{English} \\
     \cline{2-5}
     & Hotels & Books & Hotels & Books \\
    \hline
    No. Of Reviews & 373,750 & 510,598 & 515,738 & 330,000 \\
    No. of Users & 30,889 & 76,530 & NA & 17,649\\
    No. of Products & 1,858 & 4993 & 1,492 &  25,172 \\
    Avg. Reviews per Product & 264 & 102 & 345.66 & 13\\
    Max. Reviews per Product & 5,793 & 5,522 & 4,789 & 667\\
    Min. Reviews per Product & 3 & 1 & 8 & 1\\
    Median Reviews per Product & 150 & 37 & 194 & 6\\
    \hline
  \end{tabular}
\end{table}

\subsection{General Observations}
\label{sec:general-observations}
A general note when comparing the statistics of the datasets is that hotels receive a larger number of reviews than books for both cultures. This can be attributed to the intellectual nature of writing a book review which makes it harder to do when compared to writing a hotel review. We also note the large difference between the median and the average for all datasets, with the average being almost double the median. This indicates the availability of outliers in the dataset where a small portion of products receive a large number of reviews (i.e., popular products). The dataset with the largest difference in this regard is the Arabic books dataset. This shows that this dataset contained a few subsets of popular books receiving a large number of reviews.
The English book reviews dataset has a broader range of products being reviewed by a smaller set of reviewers compared to the other datasets, which explains the smaller number for the median and average for this dataset. Unfortunately, we did not have control over data collection for any of the datasets, which explains these differences. One additional discrepancy was the unavailability of a User Id for the English hotel's dataset, so we could not calculate the total number of users in it. Despite these differences, the sheer size of the datasets and the availability of an accompanying user rating with each review allowed us to gain many helpful insights into how cultures provide reviews. These insights are explained in the following sections.



\subsection{Star Ratings Analysis}
\label{sec:ratings-analysis}

\begin{figure}
     \centering
     \hfill
         \includegraphics[width=0.9\linewidth,height=5cm]{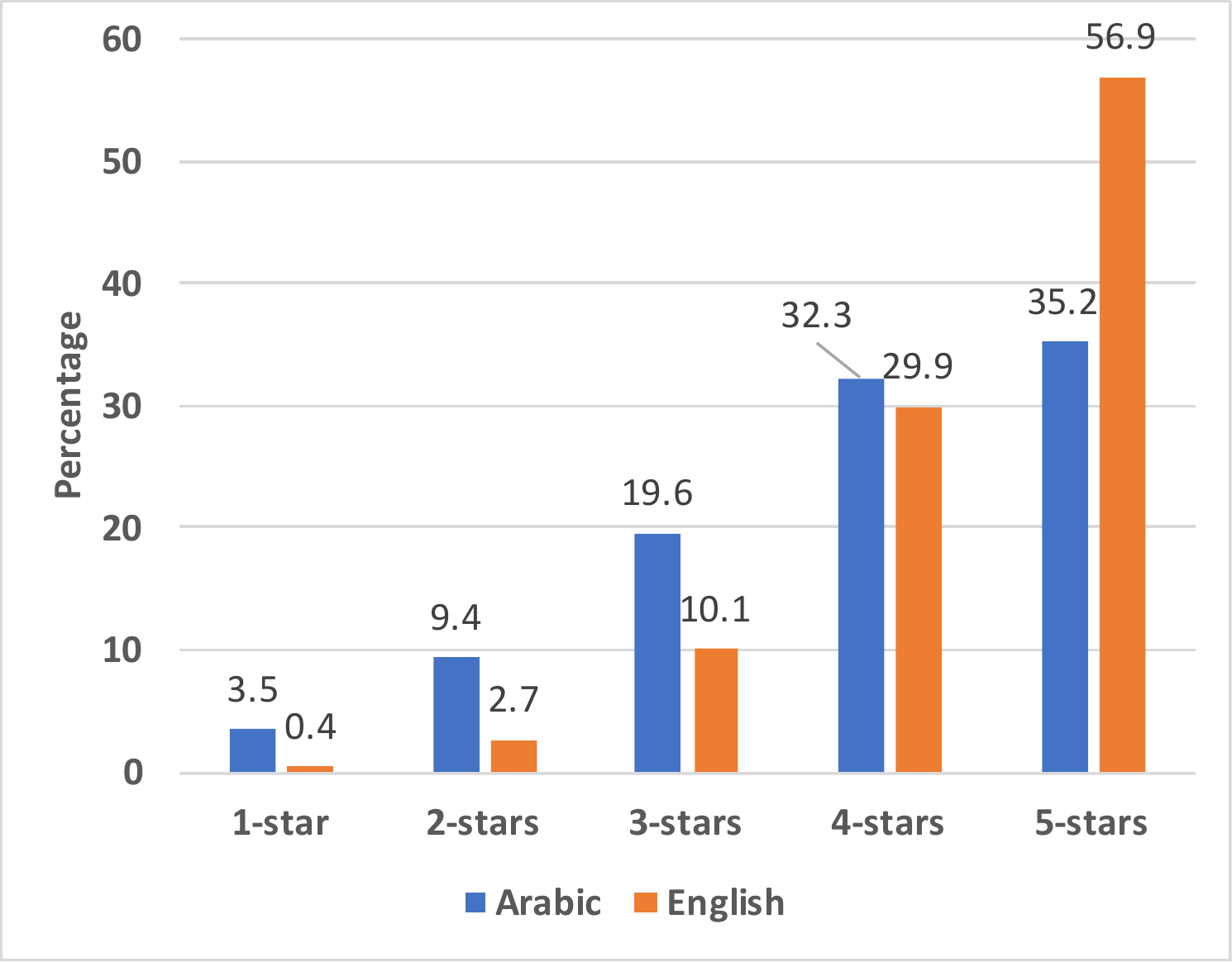}
         \caption{Star Ratings Distribution: Hotels}
         \label{fig:star-ratings-hotels}
\end{figure}

\begin{figure}
     \centering
     \hfill
         \includegraphics[width=0.9\linewidth,height=5cm]{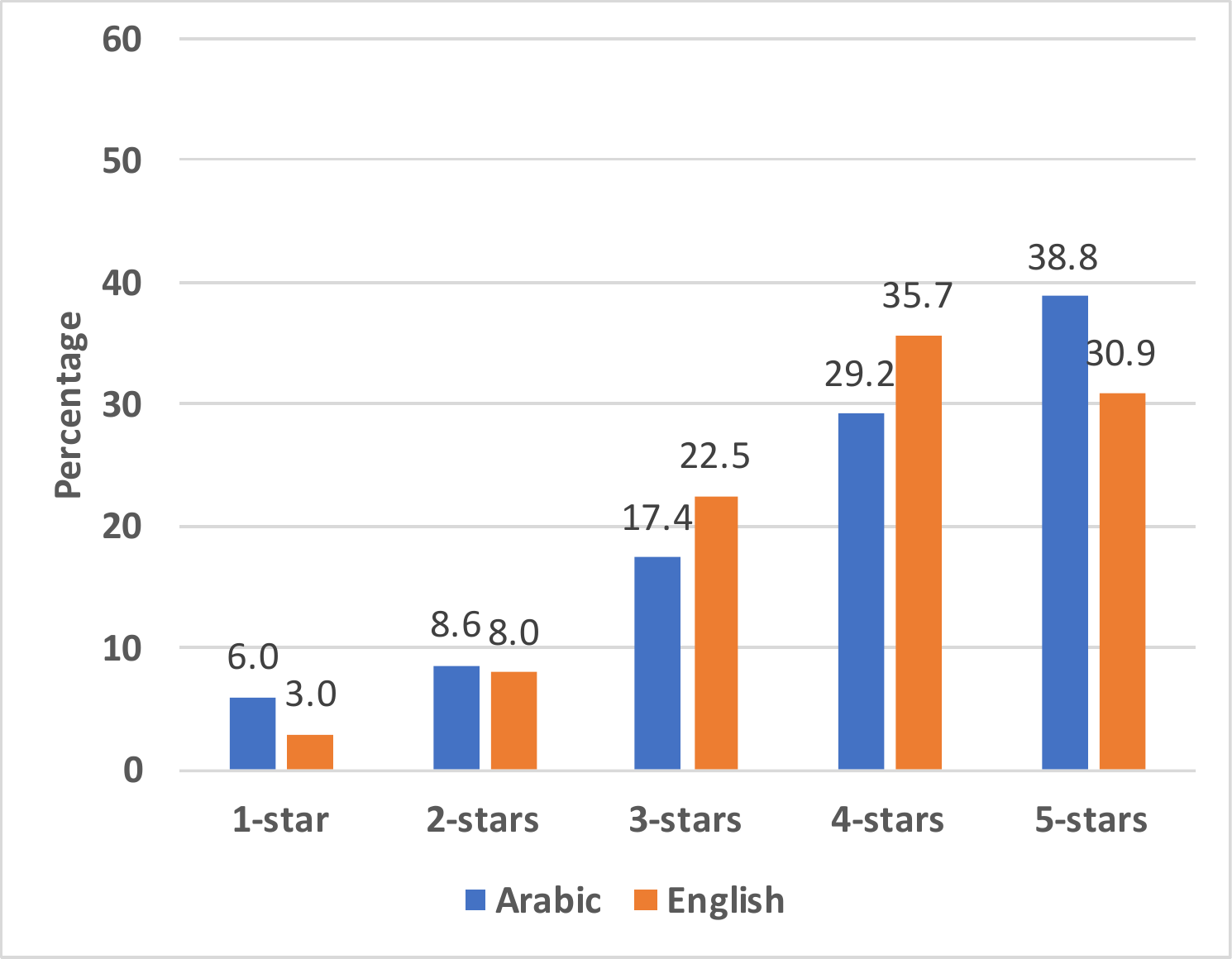}
         \caption{Star Ratings Distribution: Books}
         \label{fig:star-ratings-books}
\end{figure}

We start by analyzing how the two cultures exhibit differences when providing star ratings and whether these differences persist with different types of products. Beginning with hotels in Figure \ref{fig:star-ratings-hotels}, we see a stark difference when providing hotel star ratings between Arabic and English-speaking cultures. While English ratings are hugely skewed towards positive ratings, with 4-star and 5-star ratings constituting more than 86\% of the ratings in the dataset, the same constitutes only 67.6\% of the Arabic star ratings. This observation shows that the Arabic hotel star ratings are more distributed over the five rating levels. Also, the chances of getting a neutral star rating for Arabic reviewers is almost double the chance of getting a neutral star rating for English reviewers. The difference in distribution between star ratings for Arabic and English hotel reviews was found to be statistically significant (${\chi}^2$ = 67943, df = 4, $p<$ 0.00001). We conclude from these numbers that the Arabic culture provides more scrutiny when reviewing hotels. To validate that this trend is related to the type of product being reviewed, we plotted a similar graph comparing star ratings across the two cultures for books. When looking at Figure \ref{fig:star-ratings-books}, we see that while both English and Arabic book ratings are similarly distributed across the different star ratings, we noticed an opposite trend to what is seen in hotels where Arabic reviews contain more five-star ratings than their English counterparts, 38\% compared to 30.9\%. The difference in distribution between star ratings for Arabic and English book reviews was found to be statistically significant (${\chi}^2$ = 12682, df = 4, $p<$ 0.00001).
Nevertheless, at the negative end of the ratings, the likelihood of receiving a single star rating is still higher for Arabic reviewers than English reviewers (6\% to 3\%). It can be seen from both books and hotel reviews that the chances of getting a single-star rating are always higher for the Arabic culture. Also, when it comes to books, English reviewers provide more scrutiny; hence are more challenging to satisfy compared to Arabic reviewers, while the opposite is true for hotels. Thus both the type of product and the culture impact how users provide star ratings for products online. 

\begin{figure*}
\centering
\includegraphics[width=\linewidth,height=3cm]{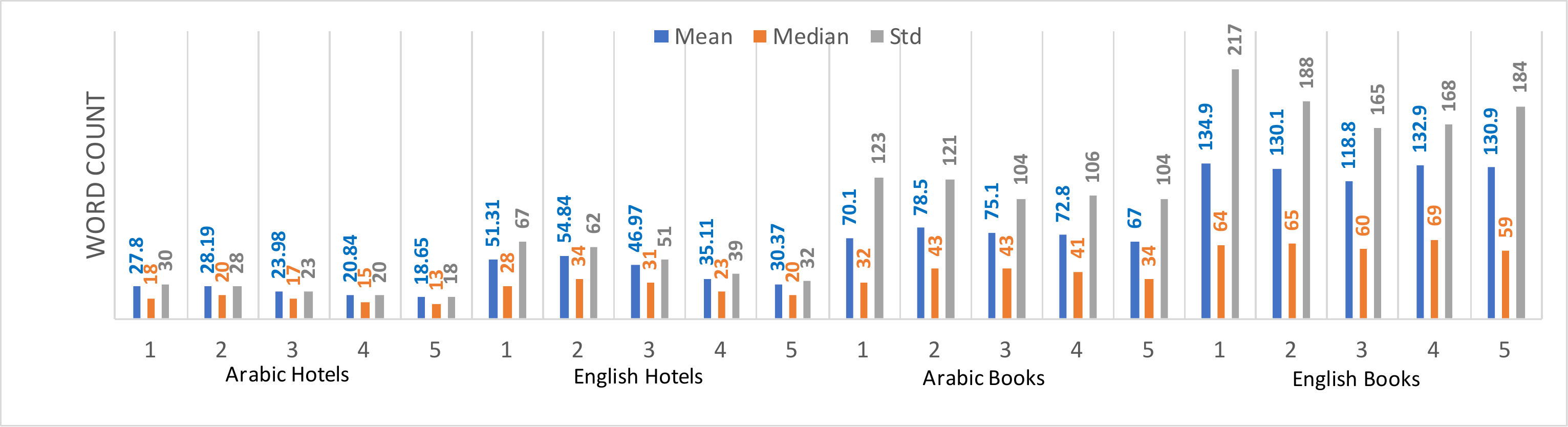}
\caption{Word Count Statistics per Star Rating for Hotels and Books datasets. }
\label{fig:word-counts-per-star-ratings}
\end{figure*}

\subsection{Review Text Analysis}
\label{sec:text-analysis}
We start the analysis for review text by inspecting differences in review length across cultures and products. Figure \ref{fig:word-counts-per-star-ratings} plots the mean, median, and standard deviation for review text word counts distributed across the five-star ratings. The distribution of review's length differs significantly among Arabic and English hotel reviews as well as Arabic and English book reviews (Mann-Whitney $U$ = 10543053927, and 1147831524, respectively, $p<$ 0.001). 
A general observation for both cultures is that book reviews have higher average word counts compared to hotels. Generally, writing a book review is an intellectual task involving details about the book, hence the higher word count. Book reviews' word counts for both cultures were double (i.e., 2x) hotel reviews' word counts. We also noticed that English reviews had higher counts than Arabic reviews, with an average increase of 25\% to 30\% in average word counts when comparing similar star ratings, which can be in part attributed to the nature of the Arabic language. Arabic is a rich language in terms of morphology where multiple words can be combined into a single word, hence the lower word count. The aforementioned two differences are related to the nature of the product and the language, which proves our analysis choice of comparing the two cultures across different products.

Now we compare the average word counts amongst the five-star ratings. Generally, negative star ratings of 1-star and 2-stars have higher average word counts than positive star ratings. However, the gap in the difference between hotels is much higher than for books for both cultures. In the case of Arabic hotels, compare the average word count of 27.8 for 1-star ratings to only 18 for the 5-star ratings, and for English hotels, compare 51.31 to 30.37. However, Arabic books compare 70.1 to 67 only and 134.9 to 130.9 for English books. When we look at the median, which is less susceptible to outliers, 4-star ratings English books received the highest word count. Also, 5-star ratings for Arabic books received more word count than 1-star ratings. This trend is not observed in hotels. Readers tend to write lengthy reviews whether they like the book or not. Whereas, the same is not valid for hotels where a lengthy review likely means that the guest may have disliked some aspect of the hotel, which led him\textbackslash her to write a lengthy review. Finally, when we compare the standard deviations across the star ratings and cultures, the gap between the standard deviation and mean is higher in books compared to hotels. We conclude from this remark that, more than hotels, books have some expert users (i.e., intellectuals) who like to write lengthy reviews about what they read. This trend is almost nonexistent for Arabic hotels where the standard deviation is close to the mean. Also, the difference between the standard deviation and the mean for English hotels is slight compared to the differences in hotel reviews.


\subsection{Review Sentiment Analysis}
\label{sec:sentiment-analysis}

\begin{figure}
\centering
\includegraphics[width=\linewidth]{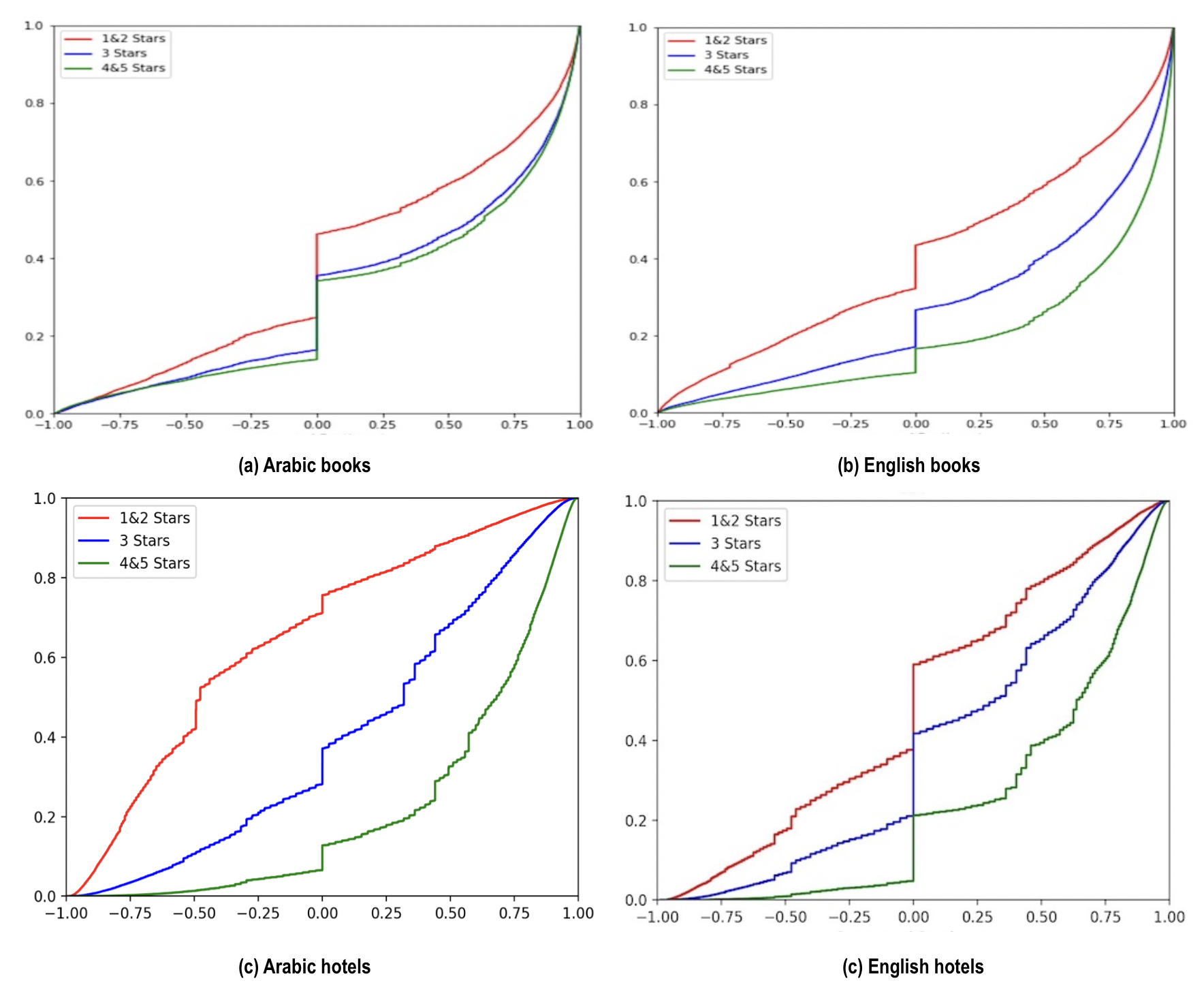}
\caption{CDFs of Compound Sentiment Score per Positive (4\&5 Stars), Negative (1\&2 Stars) and Neutral (3 Stars) Ratings.
}
\label{fig:sentiments-arabic-hotels}
\end{figure}

We now analyze the amounts of sentiments present in the text of the reviews. Figure ~\ref{fig:sentiments-arabic-hotels}  plots the CDF of the compound sentiment present in the text of reviews across positive, negative and neutral ratings. Here we combined the star ratings of 4-star and 5-star to represent positive ratings and the 1-star and 2-star ratings to represent negative ones. In contrast, a neutral rating means that the accompanying rating of the review is 3-star. We used the VADER library to generate the sentiments in the text \cite{hutto2014vader}. Since this library only works with English text, we translated the Arabic text of the Arabic datasets. We ran the library on the resultant translation to generate the sentiments for the Arabic review texts. According to other research \cite{mohammad2016translation}, the accuracy of sentiment analysis is known not to be negatively impacted by the translation. The difference in the distribution of review sentiments for Arabic and English hotel reviews was found to be statistically significant (${\chi}^2$ = 28733, df = 2, $p<$ 0.00001).

When looking at these figures,
an overall theme found for the sentiments in the four datasets is that the amount of negative sentiments for positive ratings is minimal. Conversely, negative ratings can have a significant amount of positive sentiments. When comparing products in this regard, the number of negative sentiments found in positive ratings is higher for books (Arabic books: 17\%, English books: 10\%) compared to hotels (Arabic hotels: 10\%, English hotels: 5\%). This finding indicates that if the consumer liked a hotel, they would rarely criticize it in the text. However, a book can receive more criticism despite receiving a positive rating. Also, negative ratings can contain a wealth of positive feedback that should not be ignored. We now turn into comparing cultures across products by inspecting and comparing the bottom two sub-figures in Figure \ref{fig:sentiments-arabic-hotels} that are plotting sentiments for hotel reviews. When providing a negative rating, Arabic reviewers expressed significantly more negative sentiment than English reviewers (Arabic hotels: 70\%, English hotels: 40\%). When turning into books and comparing the top two sub-figures in Figure \ref{fig:sentiments-arabic-hotels}, we can see that the amount of negative sentiments in negative ratings for the English culture across products is comparable (English hotels: 40\%, English books: 35\%). Whereas Arabic reviewers' negative sentiments were much higher for hotels (Arabic hotels: 70\%, Arabic books: 25\%). It is clear that the two cultures provide the same level of negative sentiments toward books. However, Arabic culture is critical of hotels more, especially when disliking the hotel in the rating. This finding is in harmony with the analysis of star ratings in Section \ref{sec:ratings-analysis}, which concluded that Arabic reviewers are more critical of hotels than English-speaking reviewers.

Finally, when inspecting neutral ratings, it is clear that those ratings exhibit a good amount of positive sentiments for both cultures. The chances for expressing negative sentiments toward books in neutral ratings are the same for both cultures (English books: 19\%, Arabic books: 19\%). However, the chances of receiving negative sentiments are slightly higher for Arabic hotels' neutral ratings when compared to English (English hotels: 21\%, Arabic books: 26\%).

\section{Review Helpfulness Recommendations}
\label{sec:review-helpfulness}
Understanding the cultural dimension of writing online reviews can aid in automatically processing these reviews to save users as well as business owners time from sifting through large volumes of comments \cite{zhong2018using,mankad2016understanding}. To demonstrate this, we have developed six different prediction models from review text. Three of these models provide a prediction of review helpfulness and the other three provide prediction of review star ratings, based on Arabic book review dataset, another based on English book review dataset, and the third based on the combined dataset of Arabic and English book review datasets. 
While the ground truth for predicting star rating was already available in the data sets, we needed to collect ground truth for the review helpfulness. 

\subsection{Data Labeling}
\label{subsec:Labeling}

Given that there are no review helpfulness datasets available for the Arabic language, we had to create these by manually annotating the Arabic hotel and book datasets for helpfulness of review. To do that, we created a proportionate stratified random sample of 3,000 reviews for each of the hotel and the Arabic book's datasets based on the star rating (pos, neg, neutral) and the sentiment (pos, neg, neutral). Table \ref{tab:hotel_stratisfied_sample} and Table \ref{tab:book_stratisfied_sample} show distribution of reviews per star rating and sentiment for the Arabic hotel and book samples, respectively. We uploaded these two samples to Appen \cite{Appen}, a crowdsourcing platform, to obtain labels (helpful, somewhat helpful, not helpful) for our Arabic hotel and book samples. 
{Since review helpfulness is a subjective measure that differs from user to user based on how much they find the review influence their decision making. Hence, crowdsourcing this task for getting review helpfulness labels ensures capturing this subjectivity and cultural influence on user's decision.} 

\begin{table}
  \caption{The Rating-Sentiment Distribution for the Proportionate Stratified Arabic Hotel Sample}
  \label{tab:hotel_stratisfied_sample}
  \centering
  \begin{tabular}{|l|l|}
    \hline
    strata & count \\
     \hline
pos-rating-pos-sentiment &             1762 (59\%)\\
neutral-ratings-pos-sentiment         & 358 (12\%)\\
neg-rating-neg-sentiment    &          271 (9\%)\\
neutral-ratings-neg-sentiment  &        158 (5.3\%)\\
pos-rating-neutral-sentiment &         143 (4.8\%)\\
pos-rating-neg-sentiment &             126 (4.2\%)\\
neg-rating-pos-sentiment &              90 (3\%)\\
neutral-ratings-neutral-sentiment &      67 (2.2\%)\\
neg-rating-neutral-sentiment &          25 (0.8\%)\\
  \hline
\end{tabular}
\end{table}

\begin{table}
  \caption{The Rating-Sentiment Distribution for the Proportionate Stratified Arabic Book Sample.}
  \label{tab:book_stratisfied_sample}
  \centering
  \begin{tabular}{|l|l|}
    \hline
    strata & count \\
     \hline
pos-rating-pos-sentiment &            1309 (44\%) \\

pos-rating-neutral-sentiment &         428 (14\%)
\\
neutral-ratings-pos-sentiment  &       332 (11\%)\\

pos-rating-neg-sentiment &             303 (10\%)\\

neg-rating-pos-sentiment &             262 (8.7\%)\\

neutral-ratings-neutral-sentiment&     107 (3.6\%)\\

neg-rating-neutral-sentiment&           96 (3.2\%)\\

neutral-ratings-neg-sentiment &          84 (2.8\%)\\

neg-rating-neg-sentiment &              79 (2.6\%)\\

  \hline
\end{tabular}
\end{table}

To ensure quality annotations, we created 100 test questions for each labeling task, hotels and books, to be used as a quiz in the beginning to qualify contributors and also as hidden test questions
to disqualify contributors who fall below the minimum specified accuracy. Creating test questions for such subjective tasks is not trivial. So, we created random reviews unrelated to the domain, e.g. ``we have spent a great time at this mall.". These random reviews constitute two-thirds of our test questions. The other third were reviews from the dataset itself. Contributors could choose one of four options (``helpful," ``somewhat helpful," ``not helpful," or ``unrelated"). We set the minimum accepted accuracy to be 80\% to ensure high-quality labeling. Each review got labeled by three contributors. 

Due to the scarcity of Arabic-speaking contributors on the platform, we had to suspend the tasks after waiting for about a month and not being able to get enough contributors to work on the task. However, we managed to get most of the sample data labeled with 93\% and 74\% completion rates for the hotel and the book sample, respectively. 

For each review, we considered the answer with the highest confidence score, reflecting the level of agreement among contributors weighted by their accuracy. For hotels, contributors labeled 77\% of the reviews helpful, 17.1\% somewhat helpful, 4.8\% not helpful, and 1.1\%  unrelated. For books, 72.5\% of the reviews were labeled helpful, 17.1\% somewhat helpful, 7.1\% not helpful, and 3.4\% unrelated.

In order to perform cross-cultural analysis, we needed ground truth for English reviews. Fortunately, the English book reviews dataset includes up-votes feedback for the review. In our analysis, we have considered a review to be helpful if it received three or more up-votes.

\subsection{Features}
\label{subsec:features}



\begin{table*}
\caption{Helpfulness prediction for Arabic, English and Combined books reviews}
\label{table:helpfulness-arabic}
\centering
\begin{tabular}{|c|c|c|c|c|c|c|c|c|c|c|c|c|c|c|c| } 
 \hline
 Feature & \multicolumn{5}{|c|}{Arabic} & \multicolumn{5}{|c|}{English}& \multicolumn{5}{|c|}{Combined}\\ 
 \cline{2-16}
  & Alg & Accr & Prec & Recall & F1 & Alg & Accr & Prec & Recall & F1&Alg & Accr & Prec & Recall & F1\\ 
 \hline
 Structural & NN & 0.831 & 0.776 & 0.839 & 0.806 & NN & 0.632 & 0.646 & 0.841 & 0.731 & NB & 0.760 & 0.760 & 1.000 & 0.864 \\
\hline
\textbf{TF-IDF} & NN & 0.885 & 0.846 & 0.887 & 0.866 & NN & 0.652 & 0.666 & 0.828 & 0.738 & NN & 0.802 & 0.804 & 0.978 & 0.883 \\
\hline
GALC & NN & 0.709 & 0.620 & 0.790 & 0.695 & NN & 0.638 & 0.641 & 0.887 & 0.744 & NB & 0.760 & 0.760 & 1.000 & 0.864 \\
\hline
Inquirer & RF & 0.858 & 0.815 & 0.855 & 0.835 & RF & 0.639 & 0.655 & 0.825 & 0.730 & RF & 0.765 & 0.764 & 1.000 & 0.866 \\
\hline
LIWC & NN & 0.858 & 0.836 & 0.823 & 0.829 & RF & 0.645 & 0.663 & 0.816 & 0.732 & SVM & 0.767 & 0.766 & 1.000 & 0.867 \\
\hline
\end{tabular}

\end{table*}

\begin{table*}
\caption{Star rating prediction for Arabic, English and Combined books reviews}
\label{table:star-arabic}
\centering
\begin{tabular}{|c|c|c|c|c|c|c|c|c|c|c|c|c|c|c|c| } 
 \hline
 Feature & \multicolumn{5}{|c|}{Arabic} & \multicolumn{5}{|c|}{English}& \multicolumn{5}{|c|}{Combined}\\ 
 \cline{2-16}
  & Alg & Accr & Prec & Recall & F1 & Alg & Accr & Prec & Recall & F1&Alg & Accr & Prec & Recall & F1\\ 
 \hline
Structural & NB & 0.689 & 0.689 & 1.000 & 0.816 & NN & 0.649 & 0.658 & 0.946 & 0.776 & NB & 0.645 & 0.645 & 1.000 & 0.785 \\
\hline
\textbf{TF-IDF} & NN & 0.791 & 0.838 & 0.863 & 0.850 & SVM & 0.761 & 0.742 & 0.964 & 0.839 & NN & 0.652 & 0.654 & 0.982 & 0.785 \\
\hline
GALC & DT & 0.736 & 0.730 & 0.980 & 0.837 & NN & 0.670 & 0.673 & 0.950 & 0.788 & NB & 0.645 & 0.645 & 1.000 & 0.785 \\
\hline
Inquirer & NN & 0.716 & 0.794 & 0.794 & 0.794 & SVM & 0.662 & 0.660 & 0.981 & 0.789 & NB & 0.645 & 0.645 & 1.000 & 0.785 \\
\hline
LIWC & NN & 0.703 & 0.701 & 0.990 & 0.821 & NN & 0.685 & 0.705 & 0.877 & 0.782 & RF & 0.645 & 0.645 & 1.000 & 0.785 \\
\hline
\end{tabular}

\end{table*}

\noindent \textbf{Structural Features.} We have generated review structural features used in previous research \cite{yang2015semantic} to predict review helpfulness. The features include the number of tokens, number of sentences, average sentence length, number of exclamation marks and the ratio of question sentences to the number of sentences in the review. 

\noindent \textbf{Unigram (TF-IDF).} We expected the presence of particular terms in the review to have an impact on how the reader will perceive the review as helpful or not. Hence, we generated a feature vector for each review containing the tf-idf weights for its terms unigrams. Only unigrams with document frequency greater than three were included.

\noindent \textbf{GALC.} Our analysis in Section \ref{sec:sentiment-analysis} revealed statistically significant differences in how the Arabic and English cultures express emotions in their reviews. Consequently, we decided to include GALC (Geneva Affect Label Coder) features set\cite{scherer2005emotions}. This features set is based on the number of occurrences of different words from the GALC emotion lexicon.

\noindent \textbf{Inquirer.} Inquirer features \cite{stone1962general} map each word in the review to negative, positive or neutral. We used an inquirer lexicon containing more than 11K words \cite{inquirer-file} to calculate the number of occurrences of negative and positive words in each review.

\noindent \textbf{LIWC.} We used a Linguistic Inquiry and Word Count dictionary \cite{pennebaker2001linguistic} containing 4.5K words to detect the presence of words under 55 different categories. An associated feature vector representing the number occurrences of words in each category is then constructed to be fed to our prediction models described in the next Section.

{The analysis in Section \ref{sec:analysis} revealed differences in lengths, terms used and sentiments expressed in reviews across the two cultures, which justifies our choice for the above features.} 
\subsection{Review Helpfulness Prediction}
\label{subsec:prediction}
The prediction analysis focused on books only since the English hotel reviews dataset lacked the up-votes needed to derive the helpfulness ground truth. We refrained from collecting this data using Appen due to space limitation and plan to do so as a future work. Nevertheless, the presented prediction analysis based on book reviews unveiled substantial cultural differences, which is the main scope of this work.


Feature vectors were extracted for each review text for the translated Arabic book reviews and the English book reviews. These vectors were then utilized to train six machine learning algorithms using scikit-learn python library \cite{pedregosa2011scikit} and the one that gave the best results was picked. The methods were Naive Base (NB), SVM, Decision Trees (DT), Random forests (RF), and Neural Networks (NN). The feature selection algorithm was SelectKbest with chi-square scoring function also from scikit-learn. It selects features based on the k highest scores. 

Table \ref{table:helpfulness-arabic} shows the results for helpfulness prediction for book reviews. 
Note that results in the table represent the average of 10-fold cross-validation tests.


Arabic books classification was based on 1,480 records using two labels (helpful:818 and not-helpful:662). The originally annotated dataset contained 2,404 records labeled by reviewers with four labels (helpful:1,742, not-helpful:170, somewhat-helpful:410, and unrelated:82). To build a two-way classifier, we only used the helpful labels with confidence score equal to one, meaning that all annotators agreed on the helpfulness of the review, and collapsed the three other labels under the not-helpful label, hence the reduction of record numbers from 2,404 to 1,480. On the other hand, the English books classification model was based on 10,000 records with two labels (helpful:1937, not-helpful:8063). It was generated from a larger dataset that contained 33,000 records. We selected 10,000 records using random sampling. Then, the labels were generated based on the number of votes. If the number of up-votes is greater than three, the review is helpful. Otherwise, it was considered not-helpful. For the combined result, we just merged the two datasets and performed classification based on the resultant 11,480 records. 

We can see from Table \ref{table:helpfulness-arabic} that unigram TF-IDF features with NN achieved the best accuracy for Arabic book reviews. The model predicted helpful reviews from the review text with an accuracy of 88.5\%. In the case of English books, unigram features also provided the best result however at a lower accuracy of 65.2\%. 
Since unigrams represent the words of the sentence, we conclude that the presence of certain words in each sentence was the best indicator of the review helpfulness. 
This result persisted with star ratings' prediction for Arabic and English book reviews as will be described later. The lower accuracy of the English review helpfulness prediction can be attributed to our dependence on up-votes to establish ground truth rather than collecting the data directly from the users.

We also notice that utilizing LIWC features provided equally impressive accuracy, precision, recall, and F1 score results to unigrams. Also, using structural features and inquirer feature provided good accuracy results. GALC, features provided the lowest accuracy when predicting review helpfulness for both English and Arabic book review helpfulness. 
{We also experimented with combining the features, but it did not improve prediction accuracy, and so, we do not report it here.}

Since models based on TF-IDF provide the best results, we compare the list of the most contributing terms to the model's accuracy. For Arabic books, the top five terms were {\it novel, event, writer, hero}, and {\it religion}, whereas, for English, the top five terms were {\it man, rogan, else, sexy}, and {\it laurent}. Generally speaking, this indicates that Arabic reviewers focus on specific roles while English reviewers focus on specific characters.
Another observation is that religion is an important topic for Arabic book readers, which indicates that Arab culture puts more emphasis on religion than the western culture.

Finally, for the combined English and Arabic reviews prediction the highest accuracy is still acheived using the Unigrams (TF-IDF) features. However, the accuracy went down to 80.2\% only, due to merging the less accurate English helpfulness ground truth data with the Arabic reviews data.

\subsection{Review Star Ratings Prediction}
\label{subsec:starrating}

The benefit of star ratings prediction is two-fold. First, analyzing the star rating prediction results provides insights into what makes a product (i.e. the book in our case) receive the particular star rating as will be explained in the analysis. Second, not all consumer reviews are accompanied with star ratings. Hence, such analysis can aid in automatically classifying consumer textual feedback into an individual star ratings in order for the service/product provider to focus on negative feedback for example.
The accuracy of the models can be found in Table \ref{table:star-arabic}. To generate the ground truth, we labeled the star ratings 4-star and 5-star as positive ratings, whereas 1-star, 2-star, and 3-star were considered negative ratings. The distribution for the ground truth labels was (positive: 1019, negative: 461). As for the English labels the labels were (positive: 6393, negative:3507). Similar to helpfulness models, unigram features based on TF-IDF achieved the highest accuracy. The accuracy for predicting star rating from the review text is 79.1\% and 76.1\% for Arabic and English book reviews respectively, while it goes down to 65.2\% for the combined dataset. The top five terms contributing to the Arabic book model were {\it boring, idea, wonderful, star}, and {\it naive}. On the other hand, the top five terms for the English books were 
{\it patient, acceptable, bad, disappointing}, and {\it weak}.

\subsection{Impact of Culture}
\label{subsec:impact}

To understand the impact of culture on prediction accuracy, we focus on star ratings predictions, instead of helpfulness predictions since there is a mismatch between number of up-votes and helpfulness as observed in Section~\ref{subsec:prediction}. From Table~\ref{table:star-arabic}, we see that star ratings models derived separately from Arabic books and English books provide much accuracy compared to the one derived from combined dataset. This indicates that culture plays an important role in improving the prediction accuracy. 





\section{Conclusion}
\label{sec:conclusion}

It has been estimated that 
90\% of the users read online reviews before visiting a business, and 88\% of users trust online reviews as much as personal recommendations \cite{Saleh2022}. While there is an overabundance of online reviews for every product and service, it has become difficult for the users to process and identify the helpful ones due to information overload and subsequent search and cognitive costs. Review helpfulness systems rate these reviews for helpfulness based on several factors, including message content, style, and user emotions. In this paper, we show that the cultural backgrounds of the users writing a review play a key role in assessing its helpfulness. In particular, we show the differences in the reviews between those written by Arabic users and Western users. We further show that this difference depends on the product being reviewed. Finally, we show that review helpfulness systems can provide better recommendations if the users' cultural background is used as a feature.
The next step in our work is to explore the personalization of the review recommendations based on the cultural background of the review consumers. Further, we plan to explore how some finer details such as room service or lobby area could be used to improve the review helpfulness recommendations.

\bibliographystyle{plain}
\bibliography{sample-base}

\end{document}